\documentclass[options]{JHEP3}
\title{A model for string-breaking in QCD}
\author{D. Antonov\thanks{Permanent address: ITEP, B. Cheremushkinskaya 25, RU-117 218, Moscow, Russia.}, 
L. Del Debbio, A. Di Giacomo\\  INFN-Sezione di Pisa, Universit\'a degli studi di Pisa,
Dipartimento di Fisica ``E. Fermi'', Via Buonarroti, 2 - Ed. B-C - 56127 Pisa, Italy\\
E-mail: \email{antonov, ldd, digiaco@df.unipi.it}}
\abstract{We present a model for
string breaking based on the existence of chromoelectric flux
tubes. We predict the form of the long-range potential, and obtain an
estimate of the string breaking length. A prediction is also obtained
for the behaviour with temperature of the string breaking length near the
deconfinement phase transition. We plan to use this model as a guide for a program 
of study of string breaking on the lattice.}
\preprint{IFUP-TH 2003/13}
\dedicated{}
\begin{document}


\section{Introduction}

In pure gauge theories, the potential energy of a static $\bar q q$
pair grows with the quark-antiquark distance $r$ as
\begin{equation}
V(r) = \sigma r 
\label{eq:sigma1}
\end{equation}
where $\sigma$ is known as string tension. From Regge phenomenology,
$\sigma \simeq (440\, \mathrm{ MeV})^2$. Equation~(\ref{eq:sigma1}) is
confirmed by numerical simulations on the lattice, see
e.g.~\cite{Creutz:zw}, or~\cite{Necco:2001xg} and references therein
for recent calculations.

In full QCD, due to the presence of dynamical quarks, string breaking
is expected to occur. Equation~(\ref{eq:sigma1}) yields an accurate
description of the potential up to some distance, above which the system
prefers to convert the energy into $\bar q q$ pairs, or mesons. String
breaking has been observed on the lattice, see
e.g.~\cite{Drummond:1998ir,Drummond:1998eh,Duncan:2000kr,Bernard:2001tz},
and its behaviour at finite temperature around the deconfining phase
transition has been studied numerically~\cite{DeTar:1998qa}.

In this paper, we construct a dynamical model of string
breaking. Lattice simulations have shown that the chromoelectric field
in the presence of a static quark-antiquark pair looks like an
Abrikosov flux tube~\cite{Abrikosov57} of transverse size $r_\perp$,
with $r_\perp\simeq 0.2\div 0.3\,
\mathrm{fm}$~\cite{Fukugita:1983du,Wosiek:1987kx,DiGiacomo:1990hc,Bali:1994de,Cea:1995zt,Haymaker:fm},
as conjectured since a long time~\cite{Nielsen:cs}. This picture is
consistent with a dual superconductor mechanism of
confinement~\cite{tHooft75,Mandelstam76}, the superconductor being at
the border between type-I and type-II. We shall interpret string
breaking as due to the production of light quark-antiquark pairs by
the chromoelectric field of the dual Abrikosov-Nielsen-Olesen flux tube. We plan to verify this idea by an extended program of lattice 
simulations. In this paper, we discuss the general features of the model and present some analytic calculations
in limiting cases.

A similar phenomenon occurs in electrodynamics when
one considers the potential inside a condenser. When the distance $r$
between the plates is increased, at constant electric field, the
energy increases linearly with $r$, up to some distance where the
system becomes unstable because of $e^+e^-$ pair production.

For a constant static electric field the rate of production of
spin-1/2 particle pairs per unit volume and unit time is given by~\cite{Schwinger}:
\begin{equation}
\label{eq:schwinger}
w_{\rm QED}= 2 \frac{(e E)^2}{(2\pi)^3} \sum\limits_{n=1}^{\infty}
\frac{e^{-\pi nm^2/eE}}{n^2}.
\end{equation}
In QCD the analog of the condenser is the flux tube, electron-positron
pairs are replaced by quark-antiquark pairs, the coupling constant $e$
by the strong coupling constant $g$, and a factor $N_cN_f$ must be added to properly take into account 
the number of degrees of freedom. Hence,

\begin{equation}
\label{wQCD}
w_{\rm QCD}\equiv w=2N_cN_f \frac{(g {\cal E})^2}{(2\pi)^3} \sum\limits_{n=1}^{\infty}
\frac{e^{-\pi nm^2/g{\cal E}}}{n^2},
\end{equation}
where ${\cal E}$ is a constant static chromoelectric field. 
However, free quarks do not exist in QCD, and therefore $m$ in this formula is some mass 
parameter to be determined, which can be either the constituent quark mass, or a 
typical hadron mass (e.g., $m_\pi$). On the lattice, quark masses can be changed at will, and $m$
can be understood as a function of them.

In this work, 
in order to understand the relevant physics,
we shall study the simplified case, in which the chromoelectric field is averaged over the transverse direction and then 
considered as uniform in space.
A partial physical justification is that, due to the higher
modes of the string, the tube configuration will fluctuate with a
frequency that is much higher than the rate of pair
production. More precise calculations taking into account the transverse 
size of the flux tube and the space dependence of the electric field can be performed and will be presented elsewhere.
The use of Eq.~(\ref{eq:schwinger}), which
relies on the assumption that the field is constant in space, will in any case be
a good approximation if the transverse size of the tube is larger than
$1/m$. The volume $\Omega$ that we consider is the physical
three-dimensional volume of the flux tube, of cross section $S$,
multiplied by a time equal to the length of the tube $r$ in natural
units. This is a natural choice. Indeed, in order to measure the force
between the heavy $\bar q q$ pair, we need at least the time to
propagate the interaction between the two particles. Also this choice can be tested on the lattice.
The interaction
can be described by an elastic potential if no pair has been produced
in the meantime. Hence, the long-range part of the potential is given
by:
\begin{equation}
V(r) = \sigma r\, e^{-r^2 S w}
\label{eq:potential}
\end{equation}
i.e. the linear potential $\sigma r$ multiplied by the probability
that no pairs are created in the volume $\Omega$. Clearly, 
$V(r)$ has a maximum at:
\begin{equation}
\displaystyle
\bar r = \frac{1}{\sqrt{2 S w}} 
\label{eq:rmax0}
\end{equation}
which yields an indication of the string breaking distance.
Models for hadronic processes based on similar ideas exist in the literature (see e.g.~\cite{Casher}).

\section{Explicit models}
We shall assume for the average chromoelectric field ${\cal E}$ the rms value
of the chromoelectric field on the cross section of the
tube:
\begin{eqnarray}
{\cal E} &=& \sqrt{\langle \mathcal E^2 \rangle}, \nonumber \\
\langle \mathcal E^2 \rangle &=& \frac1S \int d^2z\, \mathcal E^2(z).
\nonumber
\end{eqnarray}
At the border between type-I and type-II supeconductor (the so-called
Bogomol'nyi limit \cite{Bogomolnyi76}), $\langle \mathcal E^2 \rangle$
can be computed explicitely. The flux tube is the
solution~\cite{Abrikosov57} of the dual Landau-Ginzburg equation,
obtained from the Lagrangian:
\begin{equation}
{\cal L}=-\frac14 F^2+|D \varphi|^2
-V(\varphi).
\label{eq:AbHiggsLagr}
\end{equation}
Here $F_{\mu\nu}=\partial_\mu B_\nu-\partial_\nu B_\mu$,
$D_\mu\varphi=(\partial_\mu-ig_mB_\mu)\varphi$,
$V(\varphi)=\frac{\lambda}{2} \left(|\varphi|^2-v^2\right)^2$, 
$B_\mu$ is the dual gauge field, 
$g_m$ is the magnetic charge and, in the broken phase, $v\neq 0$. The
Bogomol'nyi limit corresponds to $m_H = m_V = \sqrt{2} g_m\, v$. In
this limit, the solutions to the equations of motion are known~\cite{deVega:1976mi} and the string
tension, defined as the integral of the energy density $\mathcal H$
over transverse directions, can be computed
exactly~\cite{Bogomolnyi76}:
\begin{equation}
\displaystyle
\sigma = \int d^2z\, \mathcal H(z) = \frac{\pi m_V^2}{g_m^2}.
\label{eq:sigma}
\end{equation}
Using the Dirac quantization condition, $g g_m = 2\pi$,
Eq.~(\ref{eq:sigma}) becomes $\sigma r_\perp^2 = \alpha_s$. 

Using the solution of Ref.~\cite{deVega:1976mi} for a tube carrying
one unit of flux, we compute the average of the electric field
squared:
\begin{equation}
\displaystyle 
\langle \mathcal E^2 \rangle = C_{\rm Bog}\frac{m_V^2\sigma}{\pi},
\label{eq:square}
\end{equation}
where $C_{\rm Bog}\simeq 0.18$.
The exponent in Eq.~(\ref{wQCD}) becomes then
\begin{equation}
-\ln y\equiv \frac{\pi^2 m^2}{\sqrt{0.18}gg_m \sigma},
\label{eq:y}
\end{equation}
and if $gg_m=2\pi$, as expected in a dual superconductor,
then $y\simeq 0.7$. 
The sum of the terms of order
higher than one in Eq.~(\ref{wQCD}) can be estimated to be
$\simeq y^2/4$, that for the worst case of $m=m_\pi$ 
allows one to keep only the first term in the
sum with an accuracy of $\sim 17\%$.
The coefficient $w$ can finally be written as:
\begin{equation}
\displaystyle
w = N_cN_f
\frac{\alpha_s}{\pi^2}\, \frac{0.18}{S}\, \sigma\, 
e^{-\pi m^2 / 2 \sqrt{0.18} \sigma}.
\label{eq:w}
\end{equation}

Note that the potential in this analysis is determined up to a
constant term, depending on the physics at shorter distances. Furthermore, $V(r)$ only depends
on the string tension $\sigma$ and the mass $m$. The potential
obtained by inserting Eq.~(\ref{eq:w}) into Eq.~(\ref{eq:potential}) can be
checked on the lattice; the dependence on $m^2$ can be tested by
comparing the potential while varying the bare quark mass.

Equation~(\ref{eq:rmax0}) yields the following expression for the position of the maximum:
\begin{equation}
\displaystyle
\bar r = \frac{\pi}{\sqrt{0.36N_cN_f \alpha_s
\sigma y}}. 
\label{eq:rmax}
\end{equation}
Note that since $\alpha_sN_c$ is constant at large $N_c$, $\bar r$ is $N_c$-independent in this limit. Next, 
using
$\alpha_s(1\, \mathrm{ GeV})\simeq 0.5$, $N_c=3$, $N_f=2$, one obtains from Eq.~(\ref{eq:rmax}) the following estimates 
$\bar r(m=200{\,}{\rm MeV})\simeq 2.0{\,}{\rm fm}$, $\bar r(m=300{\,}{\rm MeV})\simeq 3.3{\,}{\rm fm}$.  
There is a degree of
arbitrariness in the choice of the scale at which $\alpha_s$ should be
computed, so that the latter has to be considered as a parameter in
Eq.~(\ref{eq:rmax}). 
In practice~\cite{Drummond:1998ir,Drummond:1998eh,Duncan:2000kr,Bernard:2001tz},
string breaking is detected on the lattice as a deviation of the $\bar qq$-potential from the straight line $V=\sigma r$,
bigger than the numerical error. Let $\varepsilon$ be the value of such a relative error.
From Eq.~(\ref{eq:potential}) it follows

$$1-e^{-r_B^2Sw}\simeq\varepsilon,$$
or the breaking distance $r_B$:

$$r_B=\bar r\sqrt{2}\left|\ln(1-\varepsilon)\right|.$$
For instance, for $\varepsilon\simeq 20\%$, $r_B\simeq 0.28\bar r$.

The existence of the maximum of the potential and the dependence
of its position on $m$ can also be tested directly in numerical
simulations. In all this analysis the flux tube is assumed to be that
of the quenched theory. Production of gluon pairs is neglected,
because the threshold would be given by the lightest glueball mass,
which is much larger than the typical hadron mass. This is also the reason why at the one-loop order 
the formula~(\ref{eq:schwinger}) can be translated to QCD in the way this has been done in Eq.~(\ref{wQCD}).

The same analysis can be performed in the 
extreme type-II regime (else called the London limit), $\ln\kappa\gg 1$,
where 
$\kappa\equiv m_H/m_V$ is the Ginzburg-Landau parameter with $m_H=\sqrt{2\lambda}v$,
$m_V=\sqrt{2}g_mv$. The only difference
of this limit from the Bogomol'nyi one is that the coefficient 
$1/\sqrt{0.18}\simeq 2.36$ in Eq.~(\ref{eq:y}) should be replaced by 
$1.18\ln\kappa$.
Indeed, in the London limit Eq.~(\ref{eq:square}) is replaced by 

\begin{equation}
\displaystyle 
\langle \mathcal E^2 \rangle = \frac{C_{\rm Lond}}{\ln\kappa} \frac{m_V^2\sigma}{\pi},
\end{equation}
where $C_{\rm Lond}\simeq 0.71$.
For reasonable values of $\ln\kappa$ (for instance, $\ln\kappa\sim 4$), there is no significant change with
respect to the Bogomol'nyi case.

\section{Temperature dependence}

More information can be extracted from Eq.~(\ref{eq:rmax0}) if one
considers the long-range potential at finite temperature, as the
deconfinement temperature is approached.

If the mass in Eq.~(\ref{eq:schwinger}) is the constituent quark mass,
it stays finite at the transition, while the string tension vanishes
with a critical exponent:
\begin{equation}
\sigma \sim (1-T/T_c)^{\nu} \label{eq:sigmascal} 
\end{equation}
where an effective exponent $\nu=1/3$ can be used for a first order
phase transition. In this case, $\bar r$ behaves as $\frac{e^{{\rm const}/\sigma}}{\sqrt{\sigma}}$.
The approximation in Eq.~(\ref{eq:w})
is improved as we approach the phase transition. 

If instead the mass in Eq.~(\ref{eq:schwinger}) is the pion mass, 
it vanishes at the chiral point as:
\begin{equation}
m^2 \sim (1-T/T_c)^{\gamma},
\label{eq:mpiscal}
\end{equation}
where the exponent $\gamma$ is determined by the universality class of the transition. For instance,
if the symmetry breaking is described by an $O(4)$
RG fixed point~\cite{Rajagopal:1992qz}, $\gamma\simeq 1.44$. In this case, 
$y$ approaches one at the critical temperature, and the string
breaking distance grows as $1/\sqrt{\sigma}$ as the temperature is
increased.  One should also remark that Eq.~(\ref{eq:sigma}) implies
that the transverse size of the flux tube $r_\perp$ is inversely
proportional to $\sqrt{\sigma}$. When approaching $T_c$, $r_\perp^2$
goes large as $(1-T/T_c)^{-\nu}$. Since $\nu=1/3$, $r_\perp^2 m^2 \sim
(1-T/T_c)^{1.1}$ goes to zero, implying that the Schwinger formula can
no longer be applied when $T$ is very near $T_c$. These two scenarios
can be checked against data from lattice simulations.

A qualitative comparison with the data of Ref.~\cite{DeTar:1998qa} for
the string breaking length at finite temperature is roughly consistent
with the behaviour predicted by combining Eq.~(\ref{eq:rmax}) with
Eqs.~(\ref{eq:sigmascal}) and~(\ref{eq:mpiscal}). A more quantitative
analysis requires more precise and systematic numerical simulations.

\section{Concluding remarks}
We have developed a model for string breaking based on the existence
of chromoelectric flux tubes. The form of the long-range potential is
predicted, which leads to an estimate of the string breaking length;
the relevant parameters are the string tension and the mass $m$, which
is to be determined phenomenologically. A prediction is also obtained
for the behaviour of the string breaking length near the deconfinement
phase transition.  Our predictions will be tested against the results
of lattice simulations. In particular, the quark masses in lattice
simulations can be varied, and $m$ can be understood as a function of
them. 

\acknowledgments
D.A. is grateful to the whole staff of the Physics Department of the
University of Pisa for cordial hospitality.  The work has been partially
supported by INFN, by the INTAS grant Open Call 2000,
Project No. 110, and by MIUR-Progetto Teoria e fenomenologia delle
particelle elementari.

\end{document}